\documentclass[a4paper,fleqn,usenatbib,useAMS]{mnras}

\usepackage[T1]{fontenc}
\usepackage{ae,aecompl}
\usepackage{longtable}

\usepackage{graphicx}   % Including figure files
\usepackage{amsmath}    % Advanced maths commands
\usepackage{amssymb}    % Extra maths symbols
\usepackage{threeparttable}
\usepackage{tabularx}
\title[A global study of Type-B QPO in BHXBs]
  {A global study of Type B quasi-periodic oscillation in black hole X-ray binaries}

\author[Gao et al.]
{H.Q.~Gao$^{1, 2}$,
Liang Zhang$^{3}$, Yupeng Chen$^{1}$, Zhen Zhang$^{1,2}$, Li Chen$^{3}$\thanks{chenli@bnu.edu.cn (Li Chen)}, Shuang-Nan Zhang$^{1}$,
\newauthor Shu Zhang$^{1}$,Xiang Ma$^{1}$, Zi-Jian Li$^{1,2}$, Qing-Cui Bu$^{3}$, JinLu ~Qu$^{1}$\thanks{qujl@ihep.ac.cn (JinLu Qu)}\\
$^1$Key Laboratory for Particle Astrophysics, Institute of High
Energy Physics, CAS, 19B Yuquan Road,
Beijing, 100049, P. R. China\\
$^2$University of Chinese Academy of Sciences, 19A Yuquan Road, Beijing 100049, P.R. China\\
$^3$Department of Astronomy, Beijing Normal University, Beijing
100875, P. R. China.\\
}

\date{Accepted 2016 XX  XX. Received 2016 XX XX}

\pubyear{2016}

\begin{document}
\label{firstpage}
\pagerange{\pageref{firstpage}--\pageref{lastpage}}
\maketitle

\begin{abstract}
We performed a global study on the timing and spectral properties of
type-B quasi-periodic oscillations (QPOs) in the outbursts of black
hole X-ray binaries. The sample is built based on the observations
of {\it Rossi X-ray Timing Explorer}, via searching in the
literature in RXTE era for all the identified type-B QPOs. To
enlarge the sample, we also investigated some type-B QPOs that are
reported but not yet fully identified. Regarding to the time lag and
hard/soft flux ratio, we found that the sources with type-B QPOs
behave in two subgroups. In one subgroup, type-B QPO shows a hard
time lag that firstly decreases and then reverse into a soft time
lag along with softening of the energy spectrum. In the other
subgroup, type-B QPOs distribute only in a small region with hard
time lag and relatively soft hardness. These findings may be
understood with a diversity of the homogeneity showing up for the
hot inner flow of different sources. We confirm the universality of
a positive relation between the type-B QPO frequency and the hard
component luminosity in different sources. We explain the results by
considering that the type-B QPO photons are produced in the inner
accretion flow around the central black hole, under a local
Eddington limit. Using this relationship, we derived an mass estimation of $9.3-27.1$ $\rm{M_{\odot}}$ for the black hole
in H 1743-322.

\end{abstract}


%%%%%%%%%%%%%%%%%%%%%%%%%%%%%%%%%%%%%%%%%%%%%%%%%%

%%%%%%%%%%%%%%%%% BODY OF PAPER %%%%%%%%%%%%%%%%%%
\section{Introduction}
A black hole X-ray binary (BHXB) is a gravitationally bound system
consisting of a stellar mass black hole (BH) and a companion star.
They are observed and characterized by their X-ray band, spectral
and timing properties. Cygnus X-1, the first X-ray binary system
\citep{Webster1972} found for holding a BH, so far 24 sources have
been clarified as BHXBs, and a similar number of sources were
speculated as the possible BHXB candidates (see \citealt{Zhang2013};
\citealt{Remillard2006} for reviews). Most of BHXBs are transients,
which usually stay in quiescence and show occasional outbursts in
X-rays. An outburst usually shows evolution of different states
defined with the peculiar spectral/timing properties, which are
found to be well correlated with the outflow features observed in
forms of the jet and/or the disk wind. (see \citealt{Fender2012} for
review).

The systematic evolution of these states can be identified by their
X-ray spectral and timing properties(see \citealt{Zhang2013};
\citealt{Remillard2006} for reviews). It had been known that the
energy spectra of outbursting black holes often exhibit X-ray
composite spectra consisting of two broadband X-ray
components\citep{Tanaka1995}. A geometrically thin, optically thick
accretion disk \citep{Shakura1973} is thought to be the origin of
the soft X-ray component and this thermal disk is thought to be the
main structure in the high soft state (HSS). There is little doubt
that the hard X-ray component is produced with the process of
inverse Compton scattering, although the region for the Comptonising
is still a matter of debate. It may be a covering layer above the
disk (\citealt{Haardt1991}; \citealt{Haardt1993}), or some
structures above the black hole as a wind or a jet base
(\citealt{Markoff2005}; \citealt{Miller2006}), or outflows moving
away from the disk (\citealt{Beloborodov1999};
\citealt{Malzac2001}). In the truncated disk model
(\citealt{Esin1997}; \citealt{Done2007}) the thin disk truncates at
some radius larger than the last stable orbit and is replaced by
another kind of inner flow: a hot, optically thin accretion flow
which acts as both the Comptonising region and the jet base.

In the transition between the low hard state (LHS) and the HSS, the
hard component begins to be replaced by the soft component. Unlike
the spectral properties that the hard state evolves continuously and
smoothly to soft state in the hard intensity diagram (HID), abrupt
changes within such an evolution are clearly observed and lead to a
so-called isolate state, which can be distinguished to mark the
separation of transition/intermediate state. So in the spectral
hardness (the flux ratio between two different energy bands)
root-mean-square(rms) diagram (HRD), a line-like single path
consists of a low hard state, a Hard-Intermediate state (HIMS), a
high soft state and an isolate island. In this spectral evolution,
the soft-intermediate state (SIMS) occurs near the end of HIMS and
stands for the beginning of the HSS (see \citealt{Belloni2010} for a
review of this state classification). In the SIMS both the soft and
hard components have high luminosity. The inner disk temperature is
high and the hard component presents a steep energy spectrum. The
accretion structure is very unclear in this state.

During the outburst, low-frequency quasi-periodic oscillations
(QPOs; LFQPOs) with mHz to about 10 Hz are usually found. The {\it
Rossi X-ray Timing Explorer} (RXTE) mission provided the large
database of observations of the power density spectrum (PDS)
containing LFQPO in many BHXBs and led to a significant progress in
our knowledge on their timing properties. Now three main types of
LFQPOs (type A, B and C) that were identified in XTE J1550-564
originally (see \citealt{Wijnands1999}; \citealt{Homan2001};
\citealt{Remillard2002}) widely present in other BHXBs (see
\citealt{VDK2006}; \citealt{Belloni2010} for reviews). The type C
QPOs have a variable centroid frequency in the range of 0.1 - 15 Hz
and a quality factor Q ($\nu/FWHM$) larger than 6. The PDS has a
strong flat-top noise and the integrated fractional
root-mean-squared (rms) amplitude is larger than $10\%$. Type-B QPOs
have the frequency of about 1 - 8 Hz and Q larger than 6. The PDS
shows a weak power-law noise and the integrated fractional rms
amplitude is lower than $10\%$. Type A QPOs always follow type-B QPO
in the HID and show PDS similar to type-B QPO but with broader QPO
(Q $\leq 3$) and weaker amplitude (rms $\leq 5\%$ ). Since this
classification of LFQPO solely considers their intrinsically
differences in PDS (centroid frequency, quality factor, noise shape
and variability level), it has the advantage of being
phenomenological and model-independent and hence can be used in
different BHXBs and even in Z sources of neutron star systems for
probing their QPO analogy \citep{Casella2005}.

Most of the BHXB LFQPOs are found in the transition where the hard
component steepens in  energy spectrum and gradually give up the
dominance along with the growing up of a more luminous thermal
spectral component. In the HIMS, type C QPOs present. Their
frequencies continue to increase, then change to type-B QPOs with a
drop of the overall level of noise amplitude, when the source enters
the SIMS (\citealt{Belloni2010}; \citealt{Motta2011}). The
appearance of strong type-B QPOs is an indicator of the HIMS-SIMS
transition \citep{Belloni2010} and a key entrance to the HSS for an
entire `q' track outburst \citep{Zhou2013}. It is also observed with
the `jet line' and may be associated with the `moment of jet launch'
(\citealt{Soleri2008}; \citealt{Fender2009};
\citealt{Miller-Jones2012}). The type-B QPOs occur just at the
moment of the transition of X-ray energy spectral states, behaving
in phenomenon of a sharp change of timing properties and radio jet
events, which is usually believed to associate with the evolution of
the accretion flow and the launching of outflow.

In GX 339--4, \citet{Motta2011} reported type-B QPOs observed
following a different pattern in the QPO frequency versus hard flux
compared with type C and type A QPOs. \citet{Gao2014} studied the
type-B QPOs of GX 339--4 sample and suggested that they may result
from the blob related ejection event that have indeed intrinsically
different properties from the type C QPOs.
%For type C QPOs, the origin and the evolution can be explained by Lense-Thirring
%precession (\citealt{Ingram2009}; \citealt{Ingram2010}; \citealt{Ingram2011}).
In GRO J1655--40, \citet{Motta2012} reported a type-B QPO and a type
C QPO observed simultaneously in a very luminous state, suggesting
they have intrinsically different properties. In the truncated disk
scenario \citep{Done2007}, type C QPO is thought to be produced from
the Lense-Thirring procession \citep{Stella1998} of the misaligned
inner hot flow and the band-limit noise component comes from
propagation of magnetorotational instability of the same inner hot
flow. The frequency of QPO mainly relays on the truncation radius
(\citealt{Ingram2009}, \citealt{Ingram2010} and
\citealt{Ingram2011}). The peculiar properties of type-B QPOs
suggest their different origins with respect to type C QPO
(\citealt{Motta2011}, \citealt{Motta2012}). Rapid transitions among
type-B to type-A QPOs are observed in GX 339--4 \citep{Motta2011}, XTE
1817--330 \citep{Sriram2012} and XTE J1859+226 \citep{Sriram2013},
showing with evolution of the disk parameters derived from energy
spectral fittings \citep{Sriram2013}. These results suggest that
type-B and A QPOs may originally share more similarity than type-B
and C QPOs. In GX~339--4, the type-B QPO appearance in the SIMS is
always accomplished with a peak in count rate \citep{Motta2011} and
a flare in hard component flux, indicating the source spectral state
returning from HSS to SIMS. This result may suggest that there is a
sudden input of seed photons for up-compton scattering in the hot
flow \citep{Gao2014}. These abrupt seed photons may reform a small
isotropic and spherical sharp inner hot flow structure, after the
previous inner hot flow consumption in the LHS to HSS transition.
However, the underline physics corresponding to the appearance of
SIMS and the accompanied type-B QPO is not yet quite clear and is
speculated to tightly relate to the formation mechanism of type-B
QPO.

In this paper, based on the richness of the literature and the large
database of RXTE, we will try to build a sample of type-B QPOs from
all the BHTs to study the type-B QPOs with a view of global
phenomena among different BHXB systems, which have different BH
masses, distances and inclinations. In Section 2, we show the sample
selection and outline the data analysis procession. The results are
given in Section 3, and  the discussions are in Section 4.

\section[]{Data sample and analysis}

We examined all the RXTE archival observations of all the black hole
X-ray transients. For each observation, we obtained the good time
interval (GTI) using the criteria of elevation angle (ELV) $\ge$
$10^o$, offset $\le$ $0.02^o$, and the South Atlantic Anomaly (SAA)
passage time (also see \citet{Gao2014}). The observations with GTI
$\ge$ 100 s were selected as our observation sample.

\subsection{The sample}

\begin{center}
\begin{table*}

\renewcommand{\arraystretch}{1.30}
\caption{List of black hole transients with type-B QPOs} \label{log}
\begin{tabular}{ccccccc}

\hline \hline
System  &   Outbursts   &   BH Mass &   Inclination &   Distance    &   $N_{\rm H}$ &
Number of \\
        &         & ($\rm{M_{\odot}}$) &   (deg)      &   (kpc) &($\times10^{22}cm^{-2}$) &

type-B QPO Obs.\\

\hline
4U 1543--47      &           2002            &   $9.4\pm2.0$ (1)    &   $20.7\pm1.5$ (2) &   $7.5\pm0.5$ (3) &  0.43 (3)   & 3 (4)\\

GX 339--4        &   2002, 2004, 2007, 2010  &   $5.8\pm0.5$ (5)     &   $\geq40$ (6)   &   $8.0\pm4.0$ (7) &   0.5 (7) &   26 (8)\\

XTE J1752-223   &           2009            &   $9.6\pm0.9$ (9)    &   $\leq49$ (10)   & $3.5\pm0.4$ (9) &  0.46 (11)   & 1 (9)  \\

XTE J1817--330   &           2006            &   $4.0\pm2.0$ (12)    &               & $5\pm4$ (12)    &   0.15 (12)   & 8\\

XTE J1859+226   &           1999            &   $ 10.0\pm5.0$ (13)   &   $\geq60$ (14)  &   $6.2\pm1.8$ (15)&   0.34 (15) & 14 (4,16)  \\

\\

XTE J1550--564   &           1998            &   $ 10.6\pm1.0$ (2)  &  $74.7\pm3.8$ (17) &   $4.38^{+0.58}_{-0.41}$ (17) &   0.65 (18)   &   11 (4,19)\\

H 1743--322       &   2003, 2009, 2010, 2011  &   unknown             &   $75\pm3$ (20)   &   $8.5\pm0.8$ (20) &   2.4 (21) &   32 (4,22,23,24) \\

MAXI J1659--152  &           2010            &   $ 3.6-8.0$ (25)     &               & $5.3-8.6$ (25)  &   0.17 (25)   & 4 (4) \\

%GRO J1655-40 & 2005 & $7.0\pm0.2$ (16) & $3.2\pm0.2$ (17) & 0.8 (18) & 1 (19) \\
%4U 1630-47 & 1998, 2004 & [10] & $10.0\pm5.0$ (22) & 9.5 (23) & 9  \\
%XTE J1650-500 & 2001 & $6\pm3$ (29) & $2.6\pm0.7$ (30) & 0.7 (31) & 0\\
%XTE J1748-288 & 1998 & [10]  & $10\pm2 (32)$ & 7.5 (33) & 0\\
%XTE J1720-318 &  &   &  &  &0 \\
%XTE J1118+480 &  &   &  &  &0 \\
%SLX 1746-331 &  &   &  &  &0 \\
\hline \hline

\end{tabular}

%\end{center}

\begin{quote}
(1) \citet{Ritter2003}, (2) \citet{Orosz2002}, (3) \citet{Jonker2004},
(4) \citet{Motta2015},  (5) \citet{Hynes2004}, (6) \citet{Munoz2008},
(7) \citet{Zdziarski2004}, (8) \citet{Motta2011}, (9) \citet{Shaposhnikov2010},
(10) \citet{Miller-Jones2011}, (11) \citet{Markwardt2009}, (12) \citet{Sala2007},
(13) \citet{Hynes2002}, (14) \citet{Corral2011}, (15) \citet{Hynes2003},
(16) \citet{Casella2004}, (17) \citet{Orosz2011}, (18) \citet{Gierlinski2003},
(19) \citet{Remillard2002}, (20) \citet{Steiner2012}, (21) \citet{Capitanio2005},
(22) \citet{Li2013}, (23) \citet{Motta2010}, (24) \citet{Zhou2013},
(25) \citet{Yamaoka2012}.

%, (16) \citet{Hynes1998}, (17) \citet{Hjellming1995}, (18) \citet{Gierlinski2001}, (19) \citet{Motta2012}
%, (26) \citet{Orosz2004}, (27) \citet{Miniutti2004}
%, (28) \citet{Hjellming1998}, (29) \citet{Kotani2000}

\end{quote}

\end{table*}
\end{center}

In order to get a global type-B QPO, the power density spectrum was
produced for each observational sample and the type-B QPOs were
chosen following the definition of \citet{Casella2005}. The results
are then compared with the literatures (which are listed in the
Tab. \ref{log})  that reported type-B QPOs. However, \citet{Motta2015}
provided a complete LFQPO sample derived from RXTE observations that
contain 135 type-B QPOs. We use it to check our type-B QPOs sample.
Finally, in our sample, the total number of type-B QPOs is 99. The
exclusions and some notes are given in what follows.

1, The source GRS 1915+105 is excluded for its unusual outburst
evolution behavior and complex light-curve \citep{Belloni2000}.

2, We focus on the type-B QPOs in the rising phase of outbursts. The
lower frequency type-B QPO in the decay phase is excluded, which
ends up with 8 for GX 339--4 and 4 for H 1743--322.

3, GRO J1655--40 was reported with type-B QPO together with a type C
QPO in its PDS \citep{Motta2012}. There are also some other sources
holding this similar phenomenon at the peak luminosity
\citep{Li2014}. Although they share some similar properties of type
B QPO PDS, their energy spectra and states are significantly
different from SIMS type-B QPOs; we also do not take these ones into
our sample for uniformity. We also exclude those similar
observations in 4U 1630--47.

4, Some of type-B QPO in \citet{Motta2015} are inconclusive for us
which makes it hard to put them into standard type-B QPO class.
Although they are surly not type C or type A QPO, by comparing to
other normal type-B QPOs, they are remarkably distinguished by their
band-limit noise shapes.

Finally, there are eight unusual type-B QPOs in the HSS of
XTE J1817--330 during its 2006 outburst. In our sample, they are
classified into type-B QPOs because of their weak powerlaw red noise
and sharp QPO quality factor. We treat them as unusual type-B QPOs
because they also show the similarity with type A QPO in that the
rms has lower values. The HSS of this outburst is also unusual. It
was reported with a higher inner disk temperature (0.8-0.9 keV) and
a harder spectral component ($\Gamma ~$ 2.1-2.3) than other usual
HSS of BHXBs (\citealt{Roy2011}; \citealt{Sriram2012}).

\subsection{Data analysis}

\begin{figure}
\centering
     \includegraphics[width=8cm]{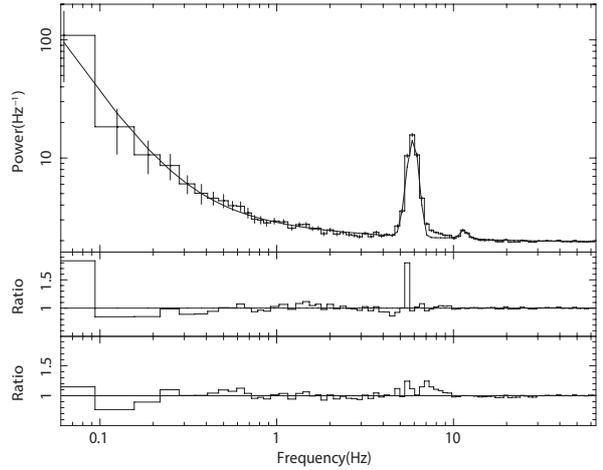}
\caption{Typical type-B QPO power density spectrum from GX 339--4. The two lower panels show the ratio residuals of models using a Lorentzian and a Gaussian to fit the QPO component, respectively.}
    \label{PDS}
\end{figure}

%(SE, SB , B and GoodXenon)
For our timing analysis, we used Event, Binned, Single-Bit and
Good Xenon data modes. The PDS was produced for each observation using POWSPEC
version 1.0 task of the XRONOS package in the channel band 0-35. We divided the data
into 64 s segments with 8 ms time resolution. The PDS was computed
for each segment and averaged with the logarithmic rebin in
frequency. The PDS was normalized using Miyamoto method (Miyamoto et
al. 1991) and the Poisson noise was subtracted. For all the type-B
QPO profile in PDS, we found that Gaussian can provide a better fitting (see Fig. \ref{PDS} ). This is also confirmed by \citet{Motta2011}. So we used a power-law component for red noise and a Gaussian component for the type QPO to get QPO
frequency, FWHM and QPO normalization in PDS from the best
fit. We calculated time/phase lags by means of
the cross-power spectrum (CPS) analysis following Qu et al. (2010),
Cui et al. (1997) and Nowak (1994). The cross-power spectrum (CPS)
is $C(j)=X^{*}_{1}(j)X_{2}(j)$, where $X_{i}(j)$ is the complex
Fourier coefficient for energy band $i$ at a given frequency
$f_{j}$. The position angle of the cross vector $C$ in the complex
plane (the Fourier phase $\phi(j)=arg[C(j)]$) are the phase lags
between the two energy bands. The time lags are $\phi(j)$ divided by
$2\pi f_{j}$. The time/phase lags of type-B QPO component were
calculated by averaging the cross vector $C(j)$ at the QPO centroid
frequency over the frequency range of the QPO FWHM. The errors were
estimated through the standard deviation of the real and imaginary
parts of the CPSs (Cui et al. 1997). We chose two broad channel
bands of 0-13 (2-5.7 keV) and 14-35 (5.7-14.8 keV) to estimate the
time/phase lags between low and high energy photons of the type-B
QPO sample.

\begin{figure}
\centering
     \includegraphics[width=8cm]{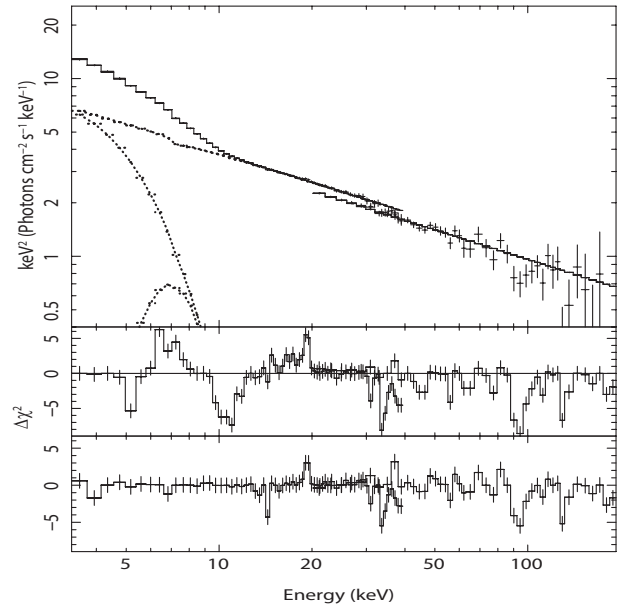}
\caption{Typical PCA and HEXTE spectra of XTE J1859+226, fitted by a model with
an absorbed disk blackbody and a powerlaw component. The two lower panels show the
reduced-$\chi^2$ without/with iron line and absorption edge respectively.}
    \label{pha}
\end{figure}

The energy spectra were extracted and fitted from background and
dead-time-corrected PCA \citep{Jahoda1996} and HEXTE \citep{Rothschild1998} data observation using FTOOLS v6.12 and XSPEC v12.7 \citep{Arnaud1996}. PCA spectra were extracted only from Standard 2 mode data born out of PCA/PCU2 and the HEXTE spectra from HEXTE/Cluster B. Considering the HEXTE/Cluster B encountered technical problems at the end of 2009, for type-B QPO observations in 2010 outburst of GX 339--4, HEXTE data were produced following \citet{Motta2011}. For the
type-B QPO observations in 2010 and 2011 outburst of H 1743--322, the
HEXTE data were not used because of the large incorrect estimation
in the background transformation form Cluster B to A. A Gaussian Fe-K emission line with centroid energies allowed vary between 6.4 and 6.7 keV was needed to obtain the best fits. The line width is constrained to a range of 0.1 -- 1.0 keV to
prevent artificial broadening due to the response of XTE/PCA at 6.4 keV \citep{Motta2011}. In addition, an iron edge component was added when its inclusion improve the fits significantly. The value of the hydrogen column
density (\emph{wabs}) for different sources was fixed (the value and
the reference in Tab. \ref{log} $N_{\rm H}$ column). We took the
model of \emph{diskbb}+\emph{powerlaw} to fit the combined PCA/HEXTE
spectra in the energy ranges 3 - 40 keV and 20 - 200 keV,
respectively and using the model of \emph{cflux} to estimate the
flux of each component in the energy ranges 2-30 keV. A systematic
error of 0.6 percent was added to account for residual uncertainties
of the instrument calibration \citep{Motta2011}. The best-fitting results have a
reduced $\chi^{2}$ between 0.85 and 1.7 in general and the error
bars are of $2.7 \sigma$ confidence level. The power-law index is in the range between 2.13 and 2.69 with an average value of $2.47\pm0.04$, which is consistent with the value in the SIMS. As an example, we show the changes of the reduced $\chi^2$ without/with Fe line and absorption edge in Fig. \ref{pha}. And we report in Tab.~\ref{tab:parameters} the frequency of the QPO, the inner disk temperature and power-law index of spectral fit for each observation.

\section{Result}

\begin{figure}
\centering
     \includegraphics[width=9cm]{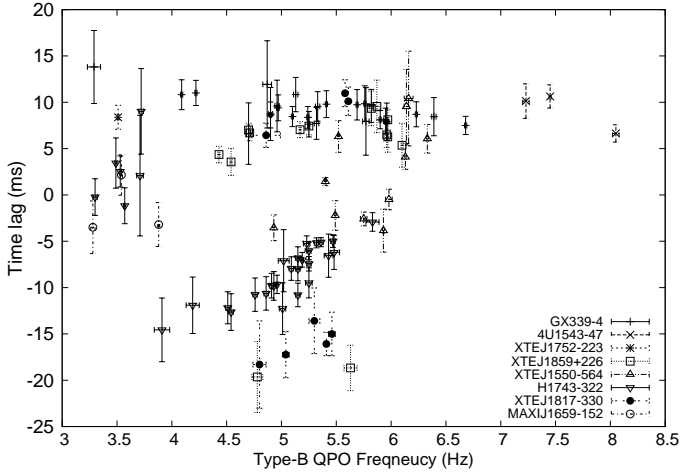}
\caption{Type-B QPO frequency versus time lag.}
    \label{fc-tlag}
\end{figure}

\begin{figure}
\centering
     \includegraphics[width=8cm]{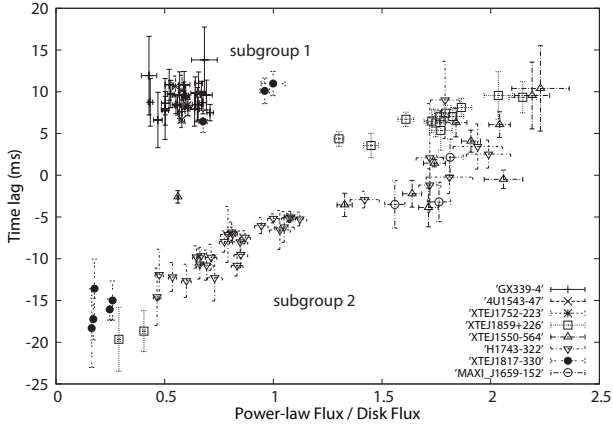}
\caption{Type-B QPO hard time lag versus the ratio of power-law flux
and disk flux.}
    \label{tlag-hardness}
\end{figure}

In Fig.~\ref{fc-tlag}, we plot the frequency of the type-B QPOs vs. the time lag. The energy band of PCA/RXTE used in time lag analysis is about 2-5.7 keV and 5.7-14.8 keV. A hard lag means that the hard photons lag the soft ones, while a soft lag means that the soft photons lag the hard ones. As shown in this figure, the time lag of the type-B QPOs distributes from $\sim10$ ms hard lag to $\sim15$ ms soft lag.  Fig.~\ref{tlag-hardness} exhibits the relation of time lag and the
`hardness ratio' ($power-law flux / disk flux$) of the type-B QPO. Two source subgroups can be distinguished from those data distribution. All the type-B QPOs of GX 339--4, 4U 1543--47 and XTE J1752--223 concentrate in the regions with a hard time lag of about 5-10 ms and a relatively small hardness ratio of about 0.4-0.7. We therefore define these sources as subgroup 1 sources. Most of type-B QPOs in this region are from the single source GX 339--4, scattering into four different outbursts. The rest of type-B QPOs are from the other sources which have a very large range
evolution in the time lag from hard time lag to soft time lag and
a large hardness ratio change. These sources enclose
H 1743--322, XTE J1859+226, XTE J1550--564, XTE J1817--330, MAXI J1659--152
and constitute subgroup 2. A clear relation established for subgroup
2 is that, along with deceasing in flux hardness ratio, the hard
time lag decreases and then changes to 15 ms soft lag finally.

\begin{table}
 \renewcommand{\tabcolsep}{0.1cm}
 \caption{Time lag spectra for six observations with type-B QPO show in Fig.~\ref{lagspectra}}
 \label{fluxfit}
 \begin{tabular}{@{}lccccc}
  \hline
&  obID   & source & QPO frequency &  time lag\\
  \hline
& 95409-01-18-05 & GX 339--4 & $4.92\pm0.02 $  & $8.89\pm2.56   $  \\
& 92082-01-02-03 & XTE J1817--330 & $4.89\pm0.01$ & $6.35\pm1.15   $ \\
& 40124-01-36-00 & XTE J1859+226 & $4.70\pm0.02 $  & $7.14\pm0.61   $  \\
& 30191-01-34-01 & XTE J1550-564 & $4.92\pm0.01 $  & $-3.93\pm1.31   $ \\
& 80146-01-56-00 & H 1743--322     & $4.85\pm0.02 $  & $-10.05\pm1.58   $ \\
& 91110-02-18-00 & XTE J1817-330 & $5.02\pm0.03$ & $-15.06\pm2.09   $ \\

\hline
 \end{tabular}
\end{table}

\begin{figure}
\centering
     \includegraphics[width=8cm]{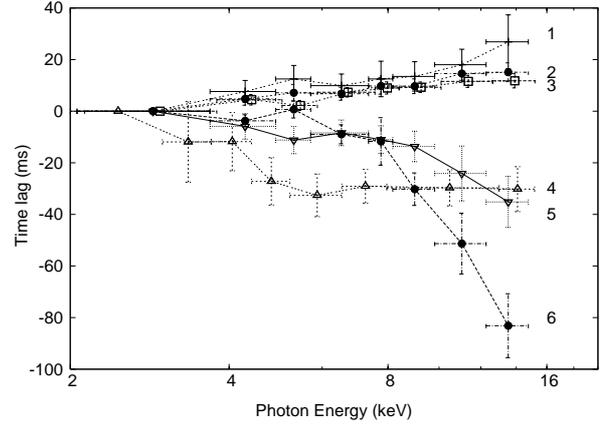}
\caption{The time lag spectra of six typical type-B QPO
observations. Points symbols are the same with other figures for
different sources.}
    \label{lagspectra}
\end{figure}

The type-B QPO observations show diverse shapes of time lag spectra.
Fig.~\ref{lagspectra} shows the time lag spectra of six typical
type-B QPO observations. We mark them with number 1 to 6 and list
the further information in Tab. \ref{fluxfit}. We chose these type-B QPOs with
the similar frequency in a narrow range of 4.7 to 5.0 Hz to minimize
the influence of frequency difference. Fig.~\ref{lagspectra} shows
that the time lag spectrum changes from number 1 to 6 following a
trend of having firstly an increasing hard time lag and then
evolving toward a decreasing soft time lag. Type-B QPO observations
of number 2 and 6 are both from XTEJ1817-330. The result shows that
they are intrinsically different in time lag spectrum shape. Type-B
QPO observation number 3 is from XTEJ1859+226. Although in
Fig.~\ref{tlag-hardness} it belongs to the subgroup 2, its time
lag spectra show no significant difference with respect to the shape
that characterizes the subgroup 1 (Number 1 of GX 339--4).

\begin{figure}
\centering
     \includegraphics[width=8cm]{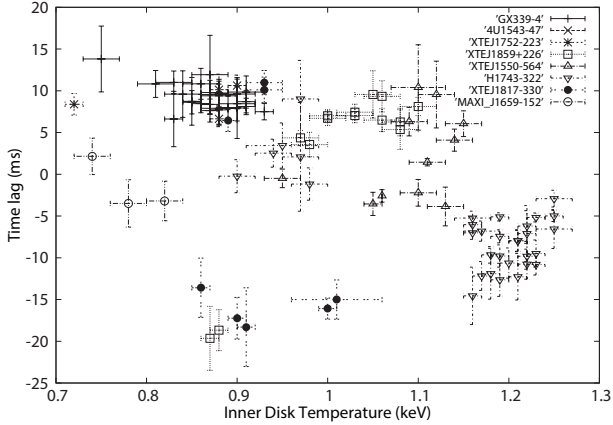}
\caption{Type-B QPO hard time lag versus inner disk temperature.}
    \label{tlag-Tin}
\end{figure}

Fig.~\ref{tlag-Tin} shows the relations between the time lag and
inner disk temperature of type-B QPOs observations. The type-B QPOs of subgroup 1 locate at the left top corner in the figure with a large hard
time lag and a low inner disk temperature. One type-B QPO
(Number 2) of XTE J1817--330 also locates in the dense region of
subgroup 1. The subgroup 2 points occupy the rest area with a larger
dispersion, especially considering  five other type-B QPOs of
XTE J1817--330. The type-B QPOs from H 1743--322 2003 outbursts
concentrate densely at the right bottom. In the middle, there exist
a few type-B QPOs from XTE J1859+226, XTE J1550--564 and six type-B
QPOs born out of outbursts of H 1743--322 during 2009-2011. There is
a negative trend between the inner disk temperature and time lag,
i.e., a hotter inner disk corresponds to a softer time lag of type-B
QPOs, if we exclude the type-B QPOs of XTE J1817--330. The source XTE
J1817--330 seems to behave in a different way in Fig. \ref{tlag-Tin}. Their time lags share two subgroups in the Fig. \ref{tlag-hardness}. The transition between hard and soft time lags suggests that there may exist a rapid change of the hot accretion
flow in a timescale of several days. But We need to verify this via further observations in future.

\begin{figure}
\centering
     \includegraphics[width=8cm]{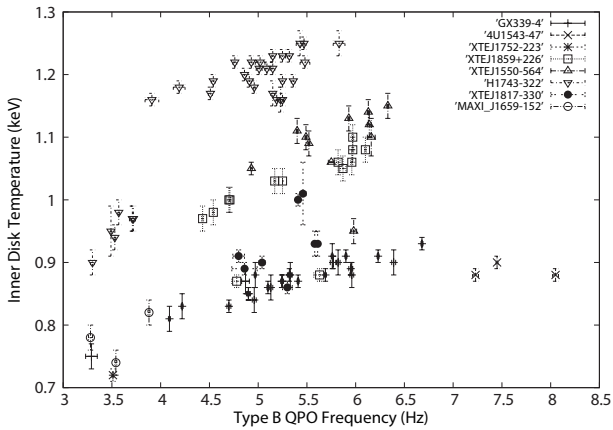}
\caption{Type-B QPO frequency versus inner disk temperature.}
    \label{Tin-fc}
\end{figure}

Fig.~\ref{Tin-fc} is the relation between the inner disk
temperature and the frequency of type-B QPOs. Different symbols
stand for the observations from different sources. The inner disk
temperature shows a tendency of increase with the frequency of type
B QPO. This relation can be found in GX339-4, XTE J1859+226,
XTEJ1550-564 and XTEJ1817-330. The highest inner disk temperature of
type-B QPO observations comes from H1743-322 (the hollow inverted
triangle points around 1.2 keV). They distribute in a relatively
crowed region and show no clear relation between inner disk
temperature and the type-B QPO frequency. But by combining the
additional six type-B QPO observations at about 0.95 keV, this
relation still holds in H1743-322. Focusing on four sources with the
largest number of SIMS type-B QPOs and the finest ABC confirmation,
the energy ranges of inner disk temperature of type-B QPO
observations are $0.99 - 1.24$ keV (H1743-322), $0.91 - 1.10$ keV
(XTEJ1550-564), $0.84 - 1.01$ keV (XTE J1859+226), $0.78 - 0.91$ keV
(GX 339--4), respectively. Within these four sources, the type-B QPOs
of GX 339--4 have the lowest inner disk temperature in general.

\begin{figure}
\centering
     \includegraphics[width=8cm]{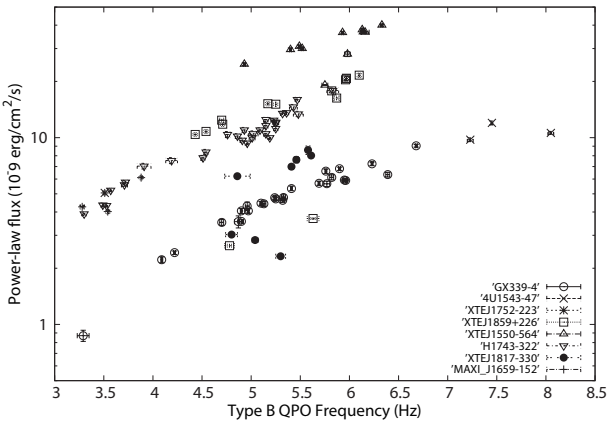}
\caption{Type-B QPO frequency versus hard component flux.}
    \label{po-fc}
\end{figure}

Fig.~\ref{po-fc} shows evolution of the type-B QPO frequency as a
function of the hard component (power-law) flux. Whichever subgroup
the source belongs to, the common positive relation between the type
B QPOs frequency and the hard component flux stand out clearly
(Fig.~\ref{po-fc}) in GX 339--4, XTE J1550--564, XTE J1859+226 and
H 1743--322.

\begin{figure*}
\centering
\includegraphics[width=0.45\linewidth]{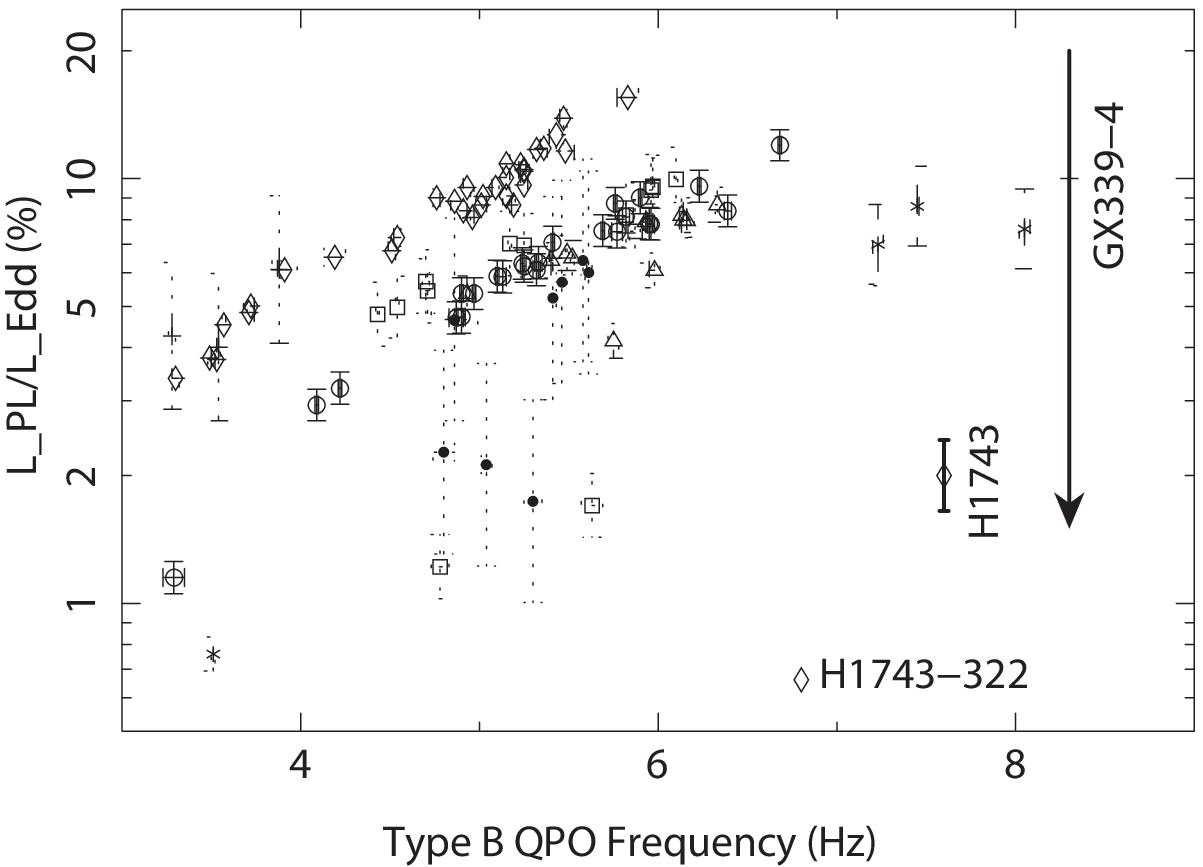}\qquad
\includegraphics[width=0.45\linewidth]{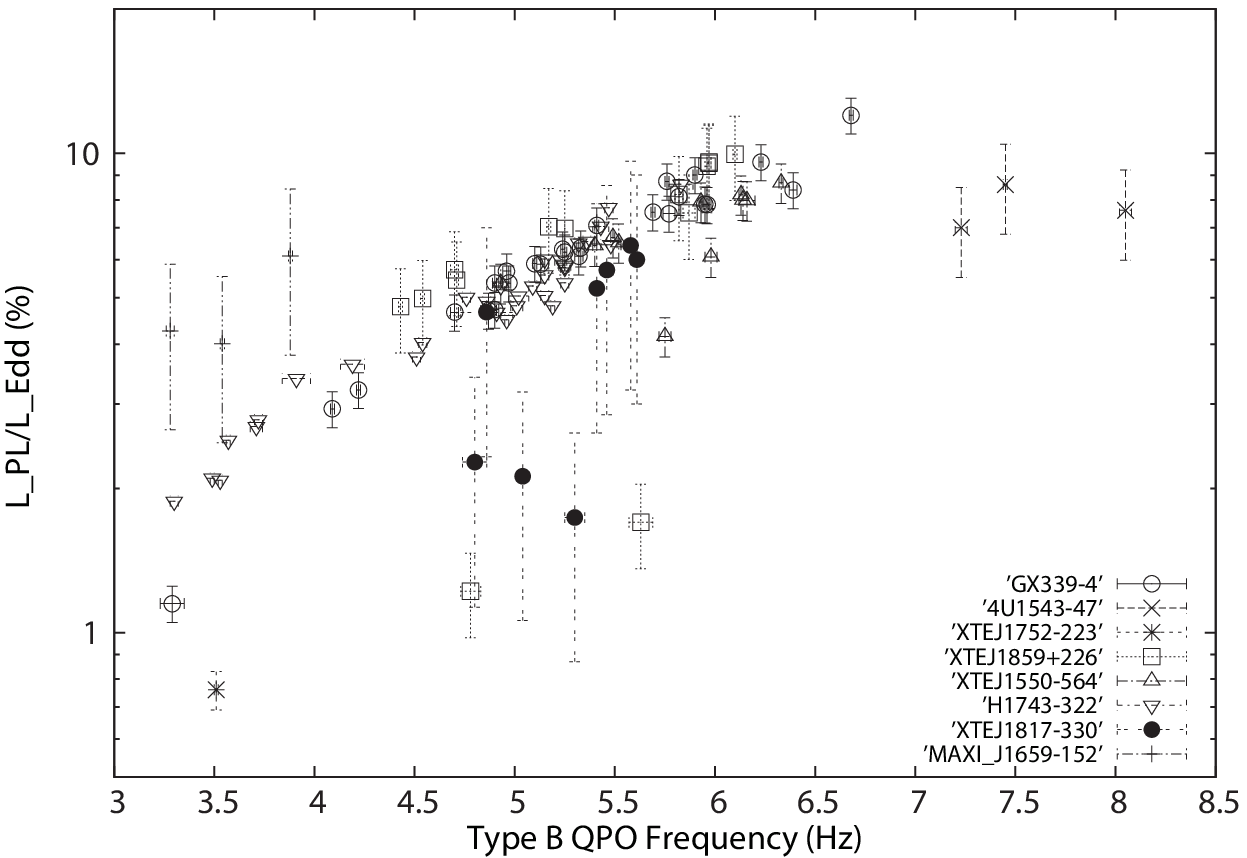}\\~\\
\caption{Hard component (Power-law) Eddington fraction versus type-B
QPO frequency. The left panel is assuming 10 solar masses for
H 1743--322. The typical distant error of GX 339--4 and H 1743--322 are
shown. Points symbols are the same with right panel. The right panel
is $18$ solar masses for H 1743--322.}
    \label{poL-fc}
\end{figure*}

Fig.~\ref{poL-fc} shows evolution of the type-B QPO frequency as a
function of the power-law luminosity (in units of Eddington
luminosity). Here, the Eddington luminosity is $L_{\rm{E}}=1.3\times10^{38}{\rm{(M/M_{\odot})}}$ erg $\rm{s}^{-1}$. The errors of the hard component Eddington luminosity mainly come from the uncertainty of the distance and black hole mass (We only show the typical error-bars caused by distance uncertainty
of the source H 1743--322 and GX~339--4 for clarity). The positive
relation of the hard component flux and frequency of type-B QPO
points from three sources (GX339-4, XTE J1859+226 and XTE J1550--564)
turns out to be in rather linear shape (solid square points in
Fig.~\ref{poL-fc}).

We should note that in Fig.~\ref{poL-fc} (left), we assumed a typical black hole mass of 10 $\rm{M_{\odot}}$ for H 1743--322 due to its mass is poorly known. The data points of H 1743--322 are located above and in parallel with the main relationship formed by other sources. We then adjusted the black hole mass of H 1743--322 to make its frequency-power-law Eddington luminosity relationship consistent with the main relationship. A black hole mass of 9.3-27.1 $\rm{M_{\odot}}$ was obtained in the 95\% confidence level.

\section{Discussion}

We discuss our results in the framework of truncated disk geometry
\citep{Done2007}: the standard disk is truncated at a radius larger
than the last stable orbit and replaced by a hot inner accretion flow.
As the accretion rate increases, the inward movement of
the truncation radius of disk corresponds to the gradual dominance of
thermal disk. It leads to the cooling, consumption and even final
collapse of the hot inner flow correlated with hard to soft state
transition. The SIMS is a very narrow period of an outburst that is
noticeable in the HID. During the transition of the LHS to HSS, the
hard component steepens and begins to drop while the soft component
becomes dominate. In the SIMS both of them are quite bright together
with the appearance of type-B QPOs \citep{Belloni2010}. They have
been found in many different sources which have different mass,
distance and inclination.

\subsection{Time lag results}

In the simple Comptonization process, the millisecond hard time lag
comes from the delay in collisions of seed soft photons with the
high energy electrons in the corona. The hard photons lag behind the
soft photons because the hard photons undergo more scatters to gain
energy (Nowak et al. 1999). It is consistent with the hard time lag
of subgroup 1 sources' type-B QPOs in Fig.~\ref{tlag-hardness}.
However if the Comptonization region is homogenous and has a uniform
temperature, no soft time lag can be produced. \citet{Nobili2000}
propose a nonhomogeneous Compton cloud model to explain hard/soft
time lag and the evolution between them with the decrease of
hardness ratio observed in state C of GRS 1919+105. The seed photons
are upscattered in the inner part region of the corona that has
higher temperature and large scattering depths than the outer part
of corona. The observed spectrum and soft lag are the effect that
the photons are downscattered in the cooler and outer part of the
corona. The relative change of disk-corona geometry and the hotter
inner part and cooler outer part of corona itself leads to the
soften of time lag and the decrease of hardness ratio. Three points
cause the observed correlation: a nonhomogeneous corona, a truncated
disk, and a source of more energetic primary photons from a smaller
inner disk radius. It is consistent with the model proposed by
\citet{Nobili2000}.

Compared with the relation of time lag and hardness ratio of state C
QPO in \citet{Nobili2000}, the similar phenomena have been observed
for subgroup 2 sources (H 1743--322, XTE J1550--564, XTE J1859+226 and
XTE J1817--330 in Fig.~\ref{tlag-hardness}) type-B QPOs in a larger
range as a whole: the time lag changes from hard to soft together
with the decrease of hardness. It suggests that they share the
similar transfer time lag progcess in a similar evolution of
accretion flow circumstance: when the inner disk edge moves inward
to the BH, the optical depth of the outer part of the nonhomogeneous
corona increases with the sharper gradient of temperature;
downscattering process becomes more important with respect to
upscattering in the corona. With the softening of the spectrum, the
hard time lag decreases and even inverses to soft. Although the
scenario of \citet{Nobili2000} was proposed for a single source, all
the type-B QPOs are in the SIMS, which is a very narrow state in the
HID and each source in subgroup 2 shows this evolution in a small
range. The results combining different sources show some kind of
large `evolution' because of the intrinsic diversity of their
accretion flow properties. The type-B QPOs of subgroup 1 sources do
not show such time lag evolution; the hard/soft flux ratio varies
slightly. It suggests that they have a more homogenous
Comptonization region and the upscattering process dominates. In
Fig. \ref{lagspectra}, all these six type-B QPO observations have
the similar QPO frequency (see Tab. \ref{fluxfit}). The significant difference
in the shape of time lag spectra of them suggest they have little
relation with the QPO frequency producing mechanism. The oscillation
photons undergo different scattering process after the type-B QPO is
produced at a certain frequency. The overall distribution of
temperature from subgroup 1 is in the left top region in
Fig. \ref{tlag-Tin}, which shows a lower inner disk temperature.
It suggests that the subgroup 1 sources have weaker ability to
provide enough energetic primary photons than subgroup 2.
Considering that the sources of subgroup 2 are high inclination
systems than subgroup1 (\citealt{Munoz2013}; \citealt{Motta2015}),
it may lead to a larger optical depth of the outer part of the
nonhomogeneous corona in the line of sight and more temperature
gradient, if the corona has a disk-like geometry. Perhaps the
inclination effect is the reason to make corona homogeneous or not
from the line of sight.

The propagating fluctuations model suggests that mass accretion rate fluctuations are generated over a wide range of radii corresponding to a wide range of time-scales. The fluctuations introduced at large radii are then propagate towards the black hole producing the observed variations in the X-ray flux \citep{Lyubarskii1997}. Since the inner regions have a harder spectrum than the outer regions, the inward propagation of fluctuations naturally give rise to hard X-ray time lags \citep{Misra00,Kotov2001}. Using this model, \citet{Arevalo06} reproduced the expected $\sim1/f$ frequency-dependence of time lags of Cyg X--1. However, \citet{Uttley2011} found that this model can explain the 2-3 vs. 0.5-0.9 keV hard lags of $\sim0.1$ s or longer, but it fails to explain the short lags (several ms) observed in our results.

Type-B QPO frequencies clearly correlate with the power-law component flux, suggesting that this type QPO are more likely to be related to the presence of a jet \citep{Motta2011}. A jet model which describes that Compton upscattering of low-energy photons in the jet can produce the hard lags \citep{Reig2003,Giannios2004,Kylafis2008}. In this model, seed photons from the accretion disk are fed into the base of the jet not into a corona, and anisotropy of the scattering process along the jet can traverse large distances in the jet, to produce large lags of longer than 0.1~s \citep{Uttley2011}. We also should note that both propagating fluctuations model and jet model are difficult to produce the soft lags observed in the type-B QPOs.

\subsection{Spectra-timing correlation results}

Among the type-B QPOs within each source, the inner disk temperature
increases with the type-B QPO frequency (Fig.~\ref{Tin-fc}). The
correlation of type-B QPOs frequencies with the power-law flux has
been observed in GX 339--4 \citep{Motta2011}. In our result, this
phenomenon has been observed in a more global sample including
GX 339--4 and both subgroup 1 and 2 sources with the same
frequency-hard component flux correlation (Fig.~\ref{po-fc}). This
common tendency in different sources suggests that the mechanism of
type-B QPO produced in the accretion flow share a common physical
process. After normalization to Eddington luminosity, a tight linear
relation among the type-B QPOs sample of GX 339--4, XTE J1859+226,
XTE J1550--564 can be found together with H 1743--322; a higher
frequency corresponds to a higher hard luminosity. It is a global
and intrinsic property of type-B QPO that may come from its special
accretion physics, which makes type-B QPO different from other type
of QPOs.

In an local Eddington-limited thin disk, the disk is in the critical
accretion state within a critical radius $r_{cr}$, and the maximum
accretion rate in the inner disk region is much less than the
classical Eddington value. The fractional critical accretion
luminosity of disk corresponds to inner disk radius: $L/L_{E}$
$\propto$ ln$\frac{r_{cr}}{r_{t}}$  (\citealt{Fukue2004};
\citealt{Heinzeller2007}; \citealt{Neilsen2011}). $L_{E}$ is the
classical Eddington luminosity. Under the assumption that type-B
QPOs are excited in the critical Eddington accretion related
process, in the Lense-Thirring QPO interpretation, an approximate
relation of the frequency and the truncation radius is $f \propto
r_{t}^{-3}$ (\citealt{Stella1998}, \citealt{VDK2006} and
\citealt{Ingram2009}). Combining them, the luminosity of the inner
disk region and the type-B QPO frequency $L/L_{E}$ $\propto$ ln$(f
\times r_{cr})$. These inner disk photons are Comptonization
upscattered in the hot inner flow. The positive relation of hard
component Eddington luminosity and the type-B QPO frequency can be
explained.

Under the assumption that the type-B QPO is produced by
Lense-Thirring mechanism in the condition of local
Eddington-limited, the intrinsic relation of type-B QPOs frequency
and hard component Eddington fractional luminosity may provide us a
new way to estimate the unknown mass of a black hole as a standard
candle. The black hole mass of H 1743--322 is poorly known. Using the
correlation between spectra index and QPO frequency, \citet{Shaposhnikov2009}
obtained a mass value of $13.3\pm3.2$.
We estimate the mass of H 1743--322 and the most consistent result
between $9.3$ and $27.1$ $\rm{M_{\odot}}$. However, we should note that
he derived mass of H1743-322 is both flux and frequency limited and may
vary for a broad range of flux and frequency. And there are some
disperse points in the Fig. \ref{poL-fc}. Our fitted results show the slopes of the linear function are consistent in the 95\% confidence level when the disperse points included or excluded. The extension of type-B QPO sample and the development of the
parameters measurement of known masses black holes will improve this
method.

\section{Summary and Conclusion}

Within the RXTE data, we selected and studied 99 type-B QPOs from
black hole X-ray binaries. We performed the timing and spectral
properties study of them and discussed the results in the framework
of truncated disk geometry.

(1) The time lag results show two subgroup behaviors. In subgroup 1,
type-B QPOs distribute only in a small region with hard time lag and
relatively soft hardness value. In subgroup 2, type-B QPOs show the
hard time lag that decreases with soften of energy spectra and
finally inverting to soft lag. The subgroup 2 sources correspond to
a nonhomogeneous corona, which is different from subgroup 1 sources.

(2) In the spectrum-timing correlation results, we confirm the
universality of the positive relation between the type-B QPOs
frequency and the hard component luminosity in different sources.
These results can be explained by that the type-B QPO photons are
produced in the condition of local Eddington critical accretion of
inner disk region. In case of having the Lense-Thirring QPO model at
work, the type-B QPO frequency and luminosity would evolve jointly
with the truncated disk radius, in a form of a positive
relationship.

(3 )Using the relation of hard component Eddington luminosity and
the type-B QPOs frequency, we estimate the unknown mass of H 1743--322
and obtain a value between $9.3$ and $27.1$ $\rm{M_{\odot}}$.

\section*{Acknowledgments}

The authors thank the RXTE team for making these data publicly
available. The authors thank anonymous referee for some helpful
suggestions and comments. This work is supported in part by the
National Natural Science Foundation of China (11173024, 11133002,
11073021, 11473027, 11103020), XTP project XDA04060604, the
Strategic Priority Research Program "The Emergence of Cosmological
Structures" of the Chinese Academy of Sciences, Grant No.
XDB09000000 and the Natural Science Foundation of China for Young
Scientists (Grant No. 11203064). The National Basic Research Program
(973 Program) of China (Grant No. 2014CB845800).

%TABLE OF PARAMETER SPACE

%%%%%%%%%%%%%%%%%%%%%%%%%%%%%%%%%%%%%%%%%%%%%%%%%%



%\bibliographystyle{mnras}
%\bibliography{BQPOs} % if your bibtex file is called example.bib

\newpage

%%_____________________BEGIN__________TABLE_1____________________________%%
%\begin{table*}
%\renewcommand{\arraystretch}{1.3}
\onecolumn

\renewcommand{\tabcolsep}{0.6cm}
\begin{center}
\begin{longtable}{|c c c c c c c|}
\caption[]{QPO and best-fit spectral parameters from all the observations on the sources of our sample.}\label{tab:parameters}\\

\endfirsthead

\multicolumn{6}{c}{{\tablename\ \thetable{} -- continued from previous page}} \\                                                                                                                                                                                                                        																											
\hline                                                                                                                                                                                                                                                                                                                                                                                                          																											
$\#$       &  ID      & Exposure & Frequency &  $kT_{\rm{in}}$ &  $\Gamma$  & ${\chi}^{2}/\nu$  \\                                                                                                                                                                                                                 																											
           &          &    (s)   &    (Hz)   &      (keV)      &            &                   \\                                                                                                                                                                                                                      																											
\hline																																
\hline	

\endhead

\hline
\multicolumn{6}{c}{{Continued on next page}} \\

\endfoot

\hline

\endlastfoot

\hline																																		
$\#$       &  ID      & Exposure & Frequency &  $kT_{\rm{in}}$ &  $\Gamma$  & ${\chi}^{2}/\nu$  \\                                                                                                                                                                                                                 																											
           &          &    (s)   &    (Hz)   &      (keV)      &            &                   \\ 																																	
\hline

\multicolumn{7}{|c|}{GX 339--4}	\\	

\hline	

1   &   60705-01-84-02  &   1680    &   $4.97\pm0.01$   &   $0.84\pm0.02$   &   $2.38\pm0.04$   &   1.10/74 \\
2   &   70108-03-02-00  &   9888    &   $5.32\pm0.02$   &   $0.83\pm0.01$   &   $2.53\pm0.03$   &   0.92/74 \\
3   &   70109-01-07-00  &   5120    &   $5.90\pm0.01$   &   $0.88\pm0.02$   &   $2.58\pm0.05$   &   0.77/74 \\
4   &   70110-01-14-00  &   880     &   $6.39\pm0.02$   &   $0.87\pm0.02$   &   $2.43\pm0.04$   &   0.98/74 \\
5   &   70110-01-15-00  &   880     &   $5.82\pm0.02$   &   $0.87\pm0.02$   &   $2.51\pm0.05$   &   1.21/74 \\
6   &   70110-01-47-00  &   1200    &   $5.77\pm0.03$   &   $0.86\pm0.02$   &   $2.57\pm0.05$   &   0.93/74 \\
7   &   90110-02-01-03  &   1056    &   $4.09\pm0.01$   &   $0.78\pm0.02$   &   $2.33\pm0.07$   &   0.82/74 \\
8   &   90704-01-02-00  &   3104    &   $4.22\pm0.01$   &   $0.78\pm0.02$   &   $2.31\pm0.04$   &   0.94/74 \\
9   &   91105-04-10-00  &   704     &   $3.29\pm0.06$   &   $0.68\pm0.04$   &   $2.22\pm0.14$   &   1.07/74 \\
10  &   92035-01-04-00  &   3232    &   $6.68\pm0.01$   &   $0.91\pm0.02$   &   $2.53\pm0.03$   &   0.44/47 \\
11  &   92085-01-03-01  &   2304    &   $6.23\pm0.01$   &   $0.87\pm0.02$   &   $2.51\pm0.03$   &   0.74/74 \\
12  &   95335-01-01-00  &   3408    &   $5.25\pm0.01$   &   $0.83\pm0.02$   &   $2.52\pm0.03$   &   0.71/74 \\
13  &   95335-01-01-01  &   2976    &   $5.10\pm0.01$   &   $0.83\pm0.02$   &   $2.53\pm0.04$   &   0.62/47 \\
14  &   95335-01-01-05  &   1520    &   $4.90\pm0.01$   &   $0.82\pm0.02$   &   $2.55\pm0.05$   &   0.87/47 \\
15  &   95335-01-01-06  &   3440    &   $4.90\pm0.02$   &   $0.81\pm0.01$   &   $2.50\pm0.04$   &   1.20/47 \\
16  &   95335-01-01-07  &   1056    &   $5.33\pm0.02$   &   $0.85\pm0.03$   &   $2.46\pm0.06$   &   0.58/47 \\
17  &   95409-01-15-02  &   1344    &   $5.76\pm0.01$   &   $0.89\pm0.02$   &   $2.52\pm0.04$   &   0.55/47 \\
18  &   95409-01-15-06  &   1424    &   $5.69\pm0.01$   &   $0.85\pm0.02$   &   $2.55\pm0.04$   &   0.68/47 \\
19  &   95409-01-16-05  &   2880    &   $5.95\pm0.02$   &   $0.86\pm0.02$   &   $2.46\pm0.03$   &   1.18/47 \\
20  &   95409-01-17-00  &   960     &   $5.96\pm0.02$   &   $0.85\pm0.02$   &   $2.45\pm0.05$   &   0.81/47 \\
21  &   95409-01-17-05  &   2464    &   $5.24\pm0.01$   &   $0.84\pm0.02$   &   $2.50\pm0.04$   &   0.81/47 \\
22  &   95409-01-17-06  &   1216    &   $5.13\pm0.02$   &   $0.83\pm0.02$   &   $2.51\pm0.05$   &   0.91/47 \\
23  &   95409-01-18-00  &   1152    &   $5.41\pm0.01$   &   $0.83\pm0.02$   &   $2.54\pm0.05$   &   0.61/47 \\
24  &   95409-01-18-04  &   144     &   $4.87\pm0.04$   &   $0.85\pm0.02$   &   $2.33\pm0.14$   &   0.57/47 \\
25  &   95409-01-18-05  &   528     &   $4.96\pm0.02$   &   $0.82\pm0.02$   &   $2.60\pm0.08$   &   0.66/47 \\
26  &   95409-01-19-00  &   896     &   $4.70\pm0.02$   &   $0.80\pm0.02$   &   $2.53\pm0.07$   &
1.04/47 \\
        \hline																									
		\hline																									
		\multicolumn{7}{|c|}{4U	1543--47}	\\																							
		\hline		
27  &   70133-01-11-00  &   1056    &   $7.45\pm0.01$   &   $0.86\pm0.02$   &   $2.67\pm0.03$   &   0.88/74 \\
28  &   70133-01-15-00  &   1184    &   $8.05\pm0.03$   &   $0.83\pm0.02$   &   $2.50\pm0.03$   &   1.05/74 \\
29  &   70133-01-16-00  &   864     &   $7.23\pm0.03$   &   $0.83\pm0.02$   &   $2.57\pm0.04$   &   0.96/74 \\
        \hline																									
		\hline																									
		\multicolumn{7}{|c|}{H 1743--322}	\\																							
		\hline	

30  &   80135-02-02-00  &   5760    &  $5.47\pm0.02$    &   $1.24\pm0.01$   &   $2.63\pm0.03$   &   0.68/74 \\
31  &   80135-02-02-000 &   15776   &   $5.32\pm0.02$   &   $1.23\pm0.01$   &   $2.68\pm0.01$   &   1.21/74 \\
32  &   80135-02-02-01  &   1424    &   $4.76\pm0.02$   &   $1.20\pm0.01$   &   $2.64\pm0.07$   &   0.95/74 \\
33  &   80144-01-01-01  &   1376    &   $5.43\pm0.04$   &   $1.24\pm0.01$   &   $2.65\pm0.06$   &   1.05/74 \\
34  &   80144-01-01-02  &   1584    &   $5.83\pm0.06$   &   $1.23\pm0.01$   &   $2.69\pm0.03$   &   0.89/74 \\
35  &   80144-01-02-00  &   1952    &   $5.25\pm0.03$   &   $1.21\pm0.02$   &   $2.61\pm0.06$   &   1.01/74 \\
36  &   80144-01-02-01  &   1408    &   $5.48\pm0.05$   &   $1.18\pm0.02$   &   $2.69\pm0.04$   &   0.68/74 \\
37  &   80144-01-03-01  &   1488    &   $5.01\pm0.03$   &   $1.18\pm0.02$   &   $2.67\pm0.04$   &   1.00/74 \\
38  &   80146-01-51-00  &   3328    &   $4.54\pm0.02$   &   $1.16\pm0.02$   &   $2.63\pm0.03$   &   0.76/74 \\
39  &   80146-01-51-01  &   3328    &   $4.19\pm0.06$   &   $1.15\pm0.02$   &   $2.58\pm0.03$   &   0.82/74 \\
40  &   80146-01-52-00  &   3344    &   $5.15\pm0.02$   &   $1.22\pm0.02$   &   $2.63\pm0.05$   &   0.88/74 \\
41  &   80146-01-52-01  &   3424    &   $5.09\pm0.02$   &   $1.19\pm0.01$   &   $2.55\pm0.05$   &   0.89/74 \\
42  &   80146-01-53-00  &   2464    &   $4.93\pm0.02$   &   $1.19\pm0.02$   &   $2.65\pm0.06$   &   1.10/74 \\
43  &   80146-01-53-01  &   1296    &   $5.02\pm0.05$   &   $1.18\pm0.03$   &   $2.67\pm0.05$   &   0.75/74 \\
44  &   80146-01-54-00  &   5440    &   $4.91\pm0.02$   &   $1.17\pm0.02$   &   $2.53\pm0.05$   &   1.13/74 \\
45  &   80146-01-55-00  &   3220    &   $5.15\pm0.02$   &   $1.18\pm0.02$   &   $2.66\pm0.03$   &   0.97/74 \\
46  &   80146-01-56-00  &   2912    &   $4.86\pm0.02$   &   $1.17\pm0.02$   &   $2.63\pm0.04$   &   0.83/74 \\
47  &   80146-01-58-00  &   3392    &   $3.91\pm0.07$   &   $1.13\pm0.02$   &   $2.55\pm0.03$   &   0.94/74 \\
48  &   80146-01-59-00  &   6560    &   $5.25\pm0.02$   &   $1.16\pm0.02$   &   $2.63\pm0.05$   &   0.84/74 \\
49  &   80146-01-60-00  &   6832    &   $5.36\pm0.02$   &   $1.17\pm0.02$   &   $2.57\pm0.04$   &   1.28/74 \\
50  &   80146-01-62-00  &   2912    &   $4.51\pm0.02$   &   $1.13\pm0.02$   &   $2.57\pm0.04$   &   0.59/74 \\
51  &   80146-01-65-00  &   4496    &   $4.96\pm0.01$   &   $1.14\pm0.02$   &   $2.60\pm0.03$   &   0.93/74 \\
52  &   80146-01-66-00  &   4160    &   $5.15\pm0.02$   &   $1.13\pm0.02$   &   $2.59\pm0.02$   &   0.79/74 \\
53  &   80146-01-67-00  &   6112    &   $5.19\pm0.01$   &   $1.13\pm0.02$   &   $2.59\pm0.04$   &   0.91/74 \\
54  &   80146-01-68-00  &   6544    &   $5.23\pm0.03$   &   $1.12\pm0.02$   &   $2.58\pm0.02$   &   0.90/74 \\
55  &   80146-01-69-00  &   4928    &   $5.25\pm0.02$   &   $1.11\pm0.02$   &   $2.59\pm0.02$   &   0.97/74 \\
56  &   94413-01-03-02  &   1392    &   $3.71\pm0.03$   &   $1.07\pm0.04$   &   $2.22\pm0.05$   &   1.06/47 \\
57  &   94413-01-03-03  &   1680    &   $3.72\pm0.02$   &   $1.08\pm0.03$   &   $2.24\pm0.04$   &   1.04/47 \\
58  &   95360-14-04-00  &   2816    &   $3.57\pm0.01$   &   $1.06\pm0.04$   &   $2.22\pm0.03$   &   1.16/47 \\
59  &   96425-01-03-00  &   1744    &   $3.53\pm0.01$   &   $1.04\pm0.04$   &   $2.19\pm0.04$   &   0.81/47 \\
60  &   96425-01-03-02  &   1344    &   $3.30\pm0.02$   &   $0.99\pm0.03$   &   $2.16\pm0.05$   &   1.10/47 \\
61  &   96425-01-03-05  &   1072    &   $3.49\pm0.01$   &   $1.03\pm0.05$   &   $2.19\pm0.05$   &   0.73/47 \\
        \hline																									
		\hline																									
		\multicolumn{7}{|c|}{XTE J1550--564}	\\																							
		\hline	
62  &   30191-01-32-00  &   880     &   $5.49\pm0.01$   &   $1.06\pm0.02$   &   $2.44\pm0.02$   &   1.05/74 \\
63  &   30191-01-33-00  &   9264    &   $5.40\pm0.01$   &   $1.05\pm0.02$   &   $2.42\pm0.01$   &   1.48/74 \\
64  &   30191-01-34-01  &   1136    &   $4.93\pm0.01$   &   $1.02\pm0.02$   &   $2.37\pm0.02$   &   1.03/74 \\
65  &   40401-01-50-00  &   3104    &   $5.75\pm0.03$   &   $1.06\pm0.01$   &   $2.37\pm0.02$   &   1.40/74 \\
66  &   40401-01-51-01  &   1120    &   $5.93\pm0.01$   &   $1.10\pm0.02$   &   $2.48\pm0.02$   &   0.78/74 \\
67  &   40401-01-53-00  &   2256    &   $6.33\pm0.01$   &   $1.10\pm0.03$   &   $2.53\pm0.02$   &   0.97/74 \\
68  &   40401-01-55-00  &   2448    &   $6.13\pm0.01$   &   $1.09\pm0.02$   &   $2.48\pm0.01$   &   1.07/74 \\
69  &   40401-01-56-00  &   544     &   $6.14\pm0.02$   &   $1.08\pm0.03$   &   $2.46\pm0.03$   &   0.86/74 \\
70  &   40401-01-56-01  &   192     &   $6.16\pm0.04$   &   $1.05\pm0.03$   &   $2.48\pm0.04$   &   1.01/74 \\
71  &   40401-01-58-00  &   1008    &   $5.52\pm0.01$   &   $1.04\pm0.02$   &   $2.39\pm0.02$   &   0.96/74 \\
72  &   40401-01-58-01  &   3456    &   $5.98\pm0.03$   &   $0.91\pm0.02$   &   $2.37\pm0.01$   &   1.12/74 \\
        \hline																									
		\hline																									
		\multicolumn{7}{|c|}{XTE J1859+226}	\\																							
		\hline	
73  &   40122-01-01-00  &   6944    &   $5.96\pm0.01$   &   $1.01\pm0.03$   &   $2.53\pm0.01$   &   0.59/74 \\
74  &   40122-01-01-02  &   2240    &   $5.97\pm0.01$   &   $1.00\pm0.03$   &   $2.53\pm0.02$   &   0.83/74 \\
75  &   40122-01-01-03  &   2496    &   $5.97\pm0.01$   &   $1.03\pm0.03$   &   $2.51\pm0.02$   &   0.91/74 \\
76  &   40124-01-13-00  &   1760    &   $6.10\pm0.01$   &   $1.03\pm0.03$   &   $2.52\pm0.02$   &   1.13/74 \\
77  &   40124-01-24-00  &   2032    &   $5.82\pm0.02$   &   $0.98\pm0.04$   &   $2.47\pm0.02$   &   0.81/74 \\
78  &   40124-01-27-00  &   1184    &   $5.87\pm0.03$   &   $0.98\pm0.04$   &   $2.44\pm0.03$   &   0.73/74 \\
79  &   40124-01-30-00  &   2624    &   $5.17\pm0.01$   &   $0.86\pm0.03$   &   $2.46\pm0.02$   &   0.88/74 \\
80  &   40124-01-36-00  &   7520    &   $4.70\pm0.02$   &   $0.94\pm0.03$   &   $2.38\pm0.02$   &   0.42/74 \\
81  &   40124-01-37-00  &   2704    &   $4.71\pm0.01$   &   $0.92\pm0.03$   &   $2.38\pm0.02$   &   0.98/74 \\
82  &   40124-01-37-01  &   2080    &   $4.54\pm0.01$   &   $0.93\pm0.03$   &   $2.35\pm0.02$   &   0.73/74 \\
83  &   40124-01-37-02  &   6288    &   $4.43\pm0.01$   &   $0.92\pm0.02$   &   $2.35\pm0.02$   &   0.51/74 \\
84  &   40124-01-39-00  &   5840    &   $5.63\pm0.01$   &   $0.86\pm0.01$   &   $2.26\pm0.03$   &   0.88/74 \\
85  &   40124-01-40-00  &   1392    &   $5.25\pm0.01$   &   $0.86\pm0.01$   &   $2.24\pm0.05$   &   0.64/74 \\
86  &   40124-01-41-00  &   2912    &   $4.78\pm0.05$   &   $0.84\pm0.01$   &   $2.23\pm0.05$   &   0.84/74 \\
        \hline																									
		\hline																									
		\multicolumn{7}{|c|}{XTE J1752--223}	\\																							
		\hline	
87  &   95360-01-01-00  &   2336    &   $3.51\pm0.01$   &   $0.68\pm0.02$   &   $2.30\pm0.03$   &   0.73/74 \\
        \hline																									
		\hline																									
		\multicolumn{7}{|c|}{XTE J1817--330}	\\																							
		\hline
88  &   91110-02-05-00  &   1728    &   $5.46\pm0.03$   &   $0.99\pm0.01$   &   $2.55\pm0.05$   &   1.32/74 \\
89  &   91110-02-07-00  &   3392    &   $5.41\pm0.02$   &   $0.98\pm0.01$   &   $2.56\pm0.05$   &   0.96/74 \\
90  &   91110-02-17-00  &   5808    &   $4.80\pm0.06$   &   $0.89\pm0.01$   &   $2.39\pm0.05$   &   0.74/74 \\
91  &   91110-02-18-00  &   7952    &   $5.04\pm0.03$   &   $0.88\pm0.01$   &   $2.36\pm0.04$   &   0.74/74 \\
92  &   91110-02-24-00  &   3888    &   $5.30\pm0.05$   &   $0.84\pm0.01$   &   $2.33\pm0.05$   &   0.74/74 \\
93  &   91110-02-29-00  &   1696    &   $5.58\pm0.02$   &   $0.87\pm0.02$   &   $2.48\pm0.03$   &   0.85/47 \\
94  &   91110-02-30-00  &   3200    &   $5.61\pm0.02$   &   $0.88\pm0.03$   &   $2.43\pm0.04$   &   0.47/47 \\
95  &   92082-01-02-03  &   1648    &   $4.86\pm0.13$   &   $0.85\pm0.02$   &   $2.45\pm0.04$   &   0.84/47 \\
        \hline																									
		\hline																									
		\multicolumn{7}{|c|}{MAXI J1659--152}	\\																							
		\hline
96  &   95118-01-01-01  &   3280    &   $3.88\pm0.01$   &   $0.89\pm0.03$   &   $2.29\pm0.03$   &   0.73/47 \\
97  &   95118-01-02-00  &   3216    &   $3.46\pm0.04$   &   $0.88\pm0.02$   &   $2.29\pm0.02$   &   0.66/47 \\
98  &   95118-01-07-00  &   3056    &   $3.28\pm0.02$   &   $0.83\pm0.03$   &   $2.19\pm0.05$   &   0.81/47 \\
99  &   95118-01-10-00  &   1088    &   $3.54\pm0.01$   &   $0.81\pm0.04$   &   $2.13\pm0.08$   &   1.32/47 \\

\hline																				
										
\end{longtable}																											
%\end{landscape}																											
\end{center}
																			
\clearpage	


%%%%%%%%%%%%%%%%%%%%%%%%%%%%%%%%%%%%%%%%%%%%%%%%%%

%%%%%%%%%%%%%%%%% APPENDICES %%%%%%%%%%%%%%%%%%%%%

%%%%%%%%%%%%%%%%%%%%%%%%%%%%%%%%%%%%%%%%%%%%%%%%%%

% Don't change these lines
\bsp    % typesetting comment
\label{lastpage}
\end{document}